\documentclass[aps,pra,longbibliography, preprint,superscriptaddress,amsmath,amssymb,floatfix]{revtex4-1}
\usepackage{graphicx}
\usepackage{dcolumn}
\usepackage{bm}
\usepackage{hyperref}

\begin{document}


\title{Hot Electrons Regain Coherence in Semiconducting Nanowires}



\author{Jonathan Reiner}
\affiliation{Department of Condensed Matter Physics, Weizmann Institute of Science, Rehovot 7610001, Israel.}

\author{Abhay Kumar Nayak}
\affiliation{Department of Condensed Matter Physics, Weizmann Institute of Science, Rehovot 7610001, Israel.}

\author{Nurit Avraham}
\affiliation{Department of Condensed Matter Physics, Weizmann Institute of Science, Rehovot 7610001, Israel.}

\author{Andrew Norris}
\affiliation{Department of Condensed Matter Physics, Weizmann Institute of Science, Rehovot 7610001, Israel.}

\author{Binghai Yan}
\affiliation{Max Planck Institute for Chemical Physics of Solids, D-01187 Dresden, Germany.}

\author{Ion Cosma Fulga}
\affiliation{Department of Condensed Matter Physics, Weizmann Institute of Science, Rehovot 7610001, Israel.}

\author{Jung-Hyun Kang}
\affiliation{Department of Condensed Matter Physics, Weizmann Institute of Science, Rehovot 7610001, Israel.}

\author{Toesten Karzig}
\affiliation{Microsoft Research, Station Q, Elings Hall, University of California, Santa Barbara, CA 93106, USA.}

\author{Hadas Shtrikman}
\affiliation{Department of Condensed Matter Physics, Weizmann Institute of Science, Rehovot 7610001, Israel.}

\author{Haim Beidenkopf}
\email[]{haim.beidenkopf@weizmann.ac.il}
\affiliation{Department of Condensed Matter Physics, Weizmann Institute of Science, Rehovot 7610001, Israel.}


\date{\today}

\begin{abstract}
The higher the energy of a particle is above equilibrium the faster it relaxes due to the growing phase-space of available electronic states it can interact with.  In the relaxation process phase coherence is lost, thus limiting high energy quantum control and manipulation. In one-dimensional systems high relaxation rates are expected to destabilize electronic quasiparticles. We show here that the decoherence induced by relaxation of hot electrons in one-dimensional semiconducting nanowires evolves non-monotonically with energy such that above a certain threshold hot-electrons regain stability with increasing energy. We directly observe this phenomenon by visualizing for the first time the interference patterns of the quasi-one-dimensional electrons using scanning tunneling microscopy. We visualize both the phase coherence length of the one-dimensional electrons, as well as their phase coherence time, captured by crystallographic Fabry-Perot resonators. A remarkable agreement with a theoretical model reveals that the non-monotonic behavior is driven by the unique manner in which one dimensional hot-electrons interact with the cold electrons occupying the Fermi-sea. This newly discovered relaxation profile suggests a high-energy regime for operating quantum applications that necessitate extended coherence or long thermalization times, and may stabilize electronic quasiparticles in one dimension.
\end{abstract}

\pacs{}

\maketitle

\section{Introduction\label{Introduction}}

The confinement of electrons to one-dimension (1D) yields a plethora of exotic phenomena. Over the years there have been various realizations of 1D electronic systems including narrow potential wells in cleaved edge overgrown tri-junctions \cite{Deshpande2010}, carbon nanotubes \cite{Deshpande2008,Ishii2003,Bockrath1999,Lee2004,Ouyang2002,Wilder1998,Odom1998}, and atomic chains\cite{Weber2012a,Nadj-Perge2014,Hager2006}. Each realization highlights different aspects of the unique nature and dynamics of electrons in 1D, altogether comprising a rich phenomenology. Among the counter-intuitive observed phenomena one finds spin-charge separation, charge fractionalization, infinite lifetime of hot-hole excitations \cite{Imambekov2012,Lunde2007,Karzig2010a,Barak2010a}, and Majorana end modes in the induced topological superconducting state of the nanowires \cite{Oreg2010,Lutchyn2010,Das2012,Mourik2012,Albrecht2016}. Here we use scanning tunneling microscopy (STM) to investigate interaction induced decoherence of 1D hot electrons confined within semiconducting InAs nanowires. We directly visualize the hot-electrons’' phase coherence length and phase coherence time \cite{Burgi1999} unveiling a novel high-energy regime of extended electronic phase coherence in 1D.

\section{Experiment\label{Experiment}}

\subsection{1D electronic structure t\label{Experiment_A}}

\begin{figure}
\includegraphics{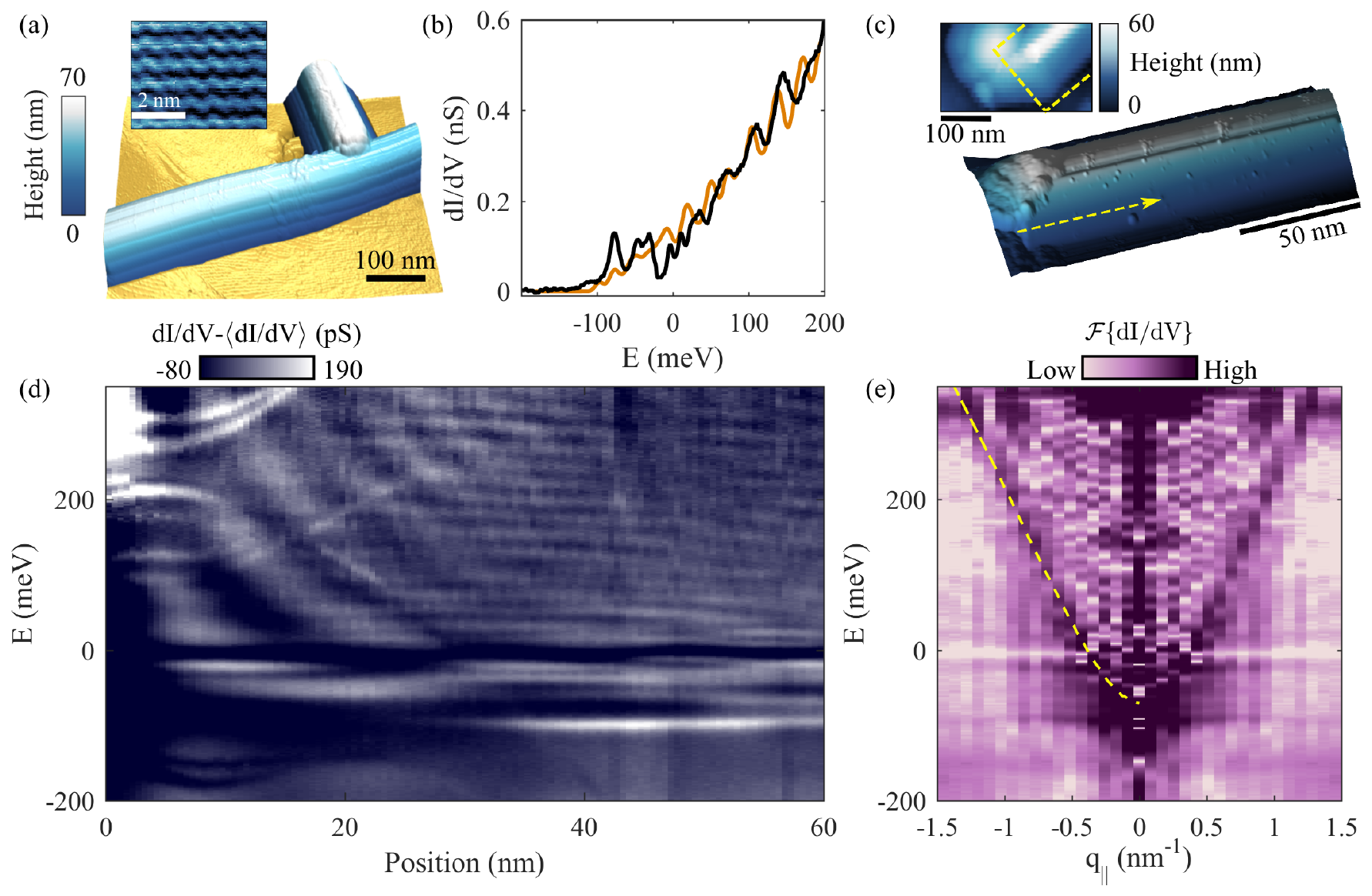} 
\caption{\label{Fig 1} Visualizing 1D subbands in semiconducting InAs nanowires. (a) STM topography of two InAs nanowires intersecting on top of the Au surface. Inset: Resolved atomic structure of the $\left \{11-20 \right \}$ facet expected of a Wurtzite nanowire grown in the $\langle 0001\rangle$ direction. (b) dI/dV measurement (black) and ab-initio calculation (orange) both showing the semiconducting gap and onset of the quantized conduction band in the form of van-Hove singularities in the LDOS. (c) Topographic image showing a clean facet at a nanowire end. Inset: topographic image of the catalyst Au droplet terminating the nanowire. The clean facet location is marked by the dashed frame. (d) dI/dV measurement along a line-cut on the clean facet (dashed arrow in (c)) showing a dominant dispersing QPI pattern above the semiconducting gap alongside non-dispersing van Hove peaks. A mean spectrum, $\langle \textrm{dI/dV} \rangle$, taken far from the nanowire end (similar to (b)) was subtracted. The residual strength of dI/dV modulations is at most 30\% of that mean value. (e) Fourier transform of (d) showing a dominant mode associated with the lowest subband of the quantized conduction band in full agreement with ab-initio calculations (dotted line). Faint dispersing QPI of higher subbands are observed as well. }
\end{figure}

In spite of the growing interest in semiconducting nanowires \cite{Ford2012,Halpern0,Holloway2015,Liang2009,Kretinin2010,Hevroni2016}, little is known on their electronic band structure and energy dispersion. In particular, very few STM spectroscopic studies have been conducted on semiconducting nanowires \cite{Hilner2008,Hjort2015,Knutsson2015,Webb2015}, although STM is potentially apt towards spectroscopic visualization on the nanoscale. The main technological challenge lies in the nanowires’ brittleness and high surface reactivity that hamper the ability to probe their surface. To overcome these challenges we have constructed a portable vacuum suitcase (Appendix  \ref{A3}) allowing transfer of the nanowires under ultra-high vacuum conditions ($< 10^{-10}$ Torr) from our RIBER molecular beam epitaxy growth chamber \cite{Kretinin2010} to our commercial (UNISOKU) STM, for subsequent measurements. The nanowires, of diameter $d=70\pm10$ nm, are harvested and distributed in situ over a freshly prepared gold (Au) substrate with a typical surface roughness of tens of nanometers, thus partially suspending the nanowires (Appendix \ref{A2}). This transfer procedure preserves the high quality of their pristine surface. Fig. \ref{Fig 1}a shows a STM topography of two intersecting nanowires lying on top of Au crests. The $\left \{11-20 \right \}$ atomic structure expected for Wurtzite nanowires grown in the $\langle 0001\rangle$ direction is clearly resolved on the pristine surface of the nanowire upper facet (inset). The presence of such a flat atomic surface, essential for STM spectroscopy, is a direct consequence of a faceting procedure added to the MBE growth protocol of the nanowires, which otherwise assume a round cross-section (see Appendix \ref{A1}).

The ability to probe the nanowire surface opens the path for in-depth study of the spectroscopic properties and dynamics of 1D hot electrons in semiconducting nanowires. The quantized nature of their energy spectrum is revealed in differential conductance (dI/dV) measurements (Fig. \ref{Fig 1}b) by a series of resonances detected above the semiconducting gap. As a property of the wavefunction confinement alone, this electronic spectrum is independent of the particular crystal termination on which it was measured,  (see Appendix \ref{A4}). The level spacings agree with a rough estimate of electrons confined to a cylinder of similar diameter. Comparison with ab-initio calculations (see Appendix \ref{B} for more details) confirms that these resonances indeed signify Van-Hove singularities in the local density of states (LDOS) located at the minima of the discrete 1D subbands that are broadened by temperature (4.2 K). The onset of the quantized conduction band falls repeatedly $\sim 100$ meV below the Fermi energy $E_{F}$ \cite{Kretinin2010}. The 1D Fermi sea of the nanowires thus occupies 3-4 1D subbands corresponding to an electron density of about 100 electrons per micron. This seems rather consistent with the amount of surface adsorbates we detect in topography. We cannot rule out, however, additional contributions of charge transfer from the Au substrate and from a surface accumulation layer \cite{Tsui1970,Olsson1996} which is thought to form in InAs nanowires.

We visualize the nanowire subbands by measuring the quasi-particle interference (QPI) patterns. These patterns are embedded in the LDOS by the 1D electrons as they scatter off crystallographic irregularities such as point impurities, stacking faults and the nanowire ends. The momentum transferred between the impinging and scattered states $q = {k_i} - {k_f}$ sets the periodicity of those modulations. Point impurities adsorbed on the surface weakly scatter the 1D electrons. Remarkably, Fourier analysis of the faint spatial modulations that emanate from randomly distributed adsorbed impurities reveals a series of parabolic bands whose minima are aligned with the spectral Van-Hove singularities (Appendix \ref{C}). To the best of our knowledge, this is the first spectroscopic visualization of the quantized subbands that form in semiconducting nanowires. 

\subsection{Phase coherence length}

To investigate the phase coherence properties of those 1D hot electrons we had to visualize the QPI patterns emanating from the strongest possible scatterer in the nanowire - its end. In STM hot electrons (holes) are injected from the tip to the sample under positive (negative) bias which allows to directly measure the extent of phase coherence of such excitations. The topographic image of the end region is shown in Fig. \ref{Fig 1}c. The dI/dV map taken across the adsorbate-free segment at the nanowire end (Fig. \ref{Fig 1}d) displays rich phenomenology including pronounced dispersing QPI patterns, spatially constant Van-Hove singularities, and evanescent resonant states that leak from the Au droplet at the nanowire end. We focus here on the origin and the nature of the pronounced dispersing QPI pattern at the nanowire end imaged in Fig. \ref{Fig 1}d. At all energies it appears as a \textit{single} dispersing wave. This is surprising in view of the numerous Van-Hove singularities we concurrently image at the same energy range, suggesting that electrons from \textit{multiple} subbands scatter from the nanowire end and interfere. Fourier analysis of the end region indeed shows a single dominant dispersing mode encompassing much fainter QPI patterns of smaller momentum transfers (Fig. \ref{Fig 1}e). The dominant 1D mode decorates the rims of the bulk conduction band in full accord with our ab-initio calculations (dashed line) and is henceforth identified with the lowest quantized subband of the conduction band. As we show in the discussion later, the enhanced intensity of the lowest subband interference pattern is a direct consequence of its significantly extended phase coherence with respect to that of higher subbands. The origin for this extended coherence length lies in the markedly restricted phase space of the relaxation processes available for hot electrons in the lowest subband with respect to those available for electrons in higher subbands.

We now take advantage of the slow relaxation of the lowest subband to visualize the energy dependence of the extended phase coherence in 1D, and in particular its high-energy revival. To quantitatively examine the phase coherence length of the lowest subband we use our precise knowledge of its wavelength dispersion (dashed line in Fig. \ref{Fig 1}e) to spatially filter out all end-features that do not follow these periodic modulations (Fig. \ref{Fig 2}a, see Appendix \ref{E1}). It thus becomes evident that the decay length of the standing wave pattern exhibits a non-monotonic behavior as a function of energy. We highlight that this non-monotonic trend appears in the raw data (Fig. \ref{Fig 1}d), and can be emphasized by other methods that eliminate non-dispersing features as spatial differentiation (Appendix \ref{E1}). The advantage of the filtering method chosen is that it allows to quantitatively extract the exponentially decaying envelope. Representative decaying standing wave patterns of the filtered data at various energies are presented in Fig. \ref{Fig 2}b. Each is fitted with an exponentially suppressed oscillating behavior, $A\left| {\sin \left( {qx + \Phi } \right)} \right|e^{-2x/L_{\varphi}}$ , where the momentum transfer q is substituted from the QPI dispersion (Fig. \ref{Fig 1}e). We stress that in the absence of interactions 1D electrons would have produced perfectly non-decaying QPI patterns. Accordingly, the exponentially decaying profiles we measure necessarily signify processes that decohere the 1D electrons. 

The energy evolution of the fitted phase coherence length $L_{\varphi}$ is presented in Fig. \ref{Fig 2}c.  In the hot-holes sector, $E<E_{F}$, we find a long coherence length of the standing wave pattern. It reaches a maximum at the Fermi energy of order $L_{\varphi}(E_{F}) \sim 100$ nm corresponding to phase coherence time of $\sim 100$ fs \cite{Liang2009}. Note, that this is not the thermalization time but rather the first relaxation step the hot electron undergoes towards it. The energy evolution  sharply changes upon crossing $E_{F}$ to the hot-electron sector, $E>E_{F}$,  where the phase coherence length starts to shorten rapidly, demonstrating the asymmetric electron-hole relaxation in 1D \cite{Barak2010a}. This initial trend seems to agree with Landau'’s principle under which the higher a hot-electron is above $E_{F}$ the broader the available shell of occupied states it interacts with  (demonstrated by the dashed line in Fig. \ref{Fig 2}c). Remarkably, however, we find that this intuitive trend of coherence loss with increasing energy reverses above $\sim 80$ meV where the ultra-hot electrons' phase coherence length begins to grow back with energy. At the highest energies measured the phase coherence length is comparable to the one we find at the Fermi energy.

\begin{figure}
\includegraphics{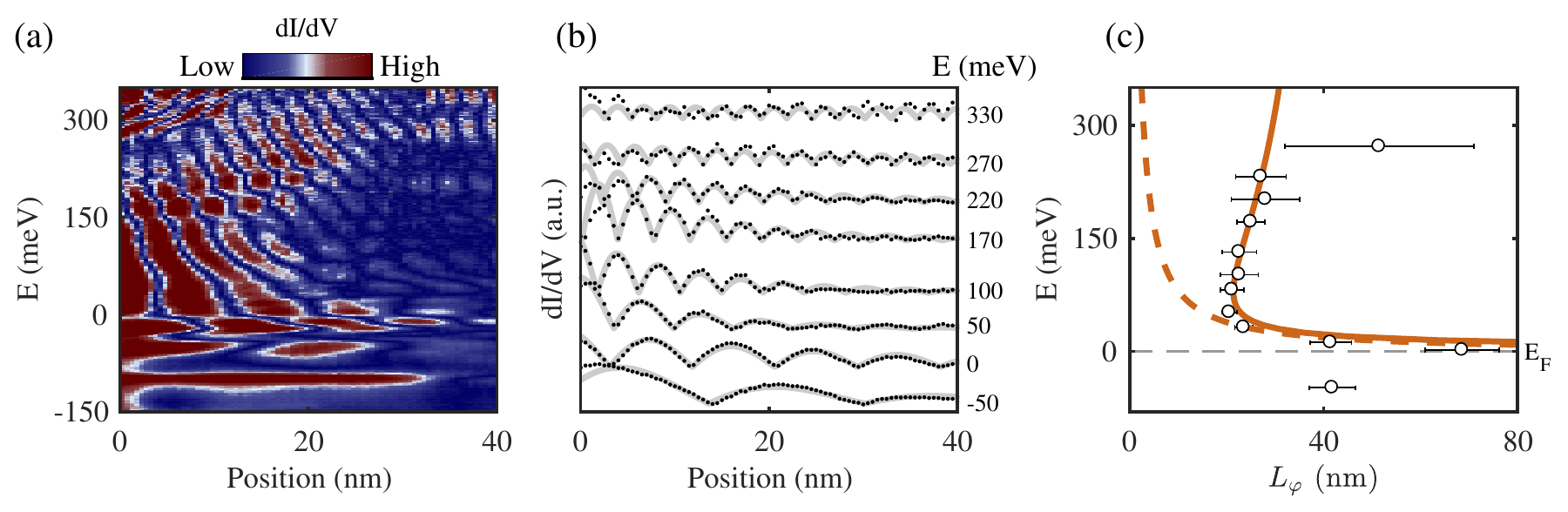} 
\caption{\label{Fig 2} Non-monotonic phase coherence length $L_{\varphi}$ visualized by QPI decay. (a) The dI/dV map from Fig. \ref{Fig 1}d after applying a '‘spatial lock-in'’ filtration (see Appendix \ref{E1}), used to highlight the decay profile of the QPI patterns. The extended standing wave patterns of the hot-holes turn into progressively decaying QPI patterns above $E_{F}$ but then revive above $\sim 80$ meV.  (b) Representative decay profiles from (a) (dotted lines), each is fitted to  $A\left| {\sin \left( {qx + \Phi } \right)} \right|e^{-2x/L_{\varphi}}$ (gray lines). At the highest measured energies the decay is too slow to fit reliably within the measured distance, which is limited by the distant surface impurities (hence the large error bars). (c) The fitted phase coherence length at different energies displays a non-monotonic evolution with energy (open circles). The calculated phase coherence length from the model (solid line) agrees well with the measured non-monotonic trend, and stands in sharp contrast to the $E^{-2}$ behavior expected at the low energy regime   (dashed line).}
\end{figure}

\subsection{Phase coherence time}

\begin{figure}
\includegraphics{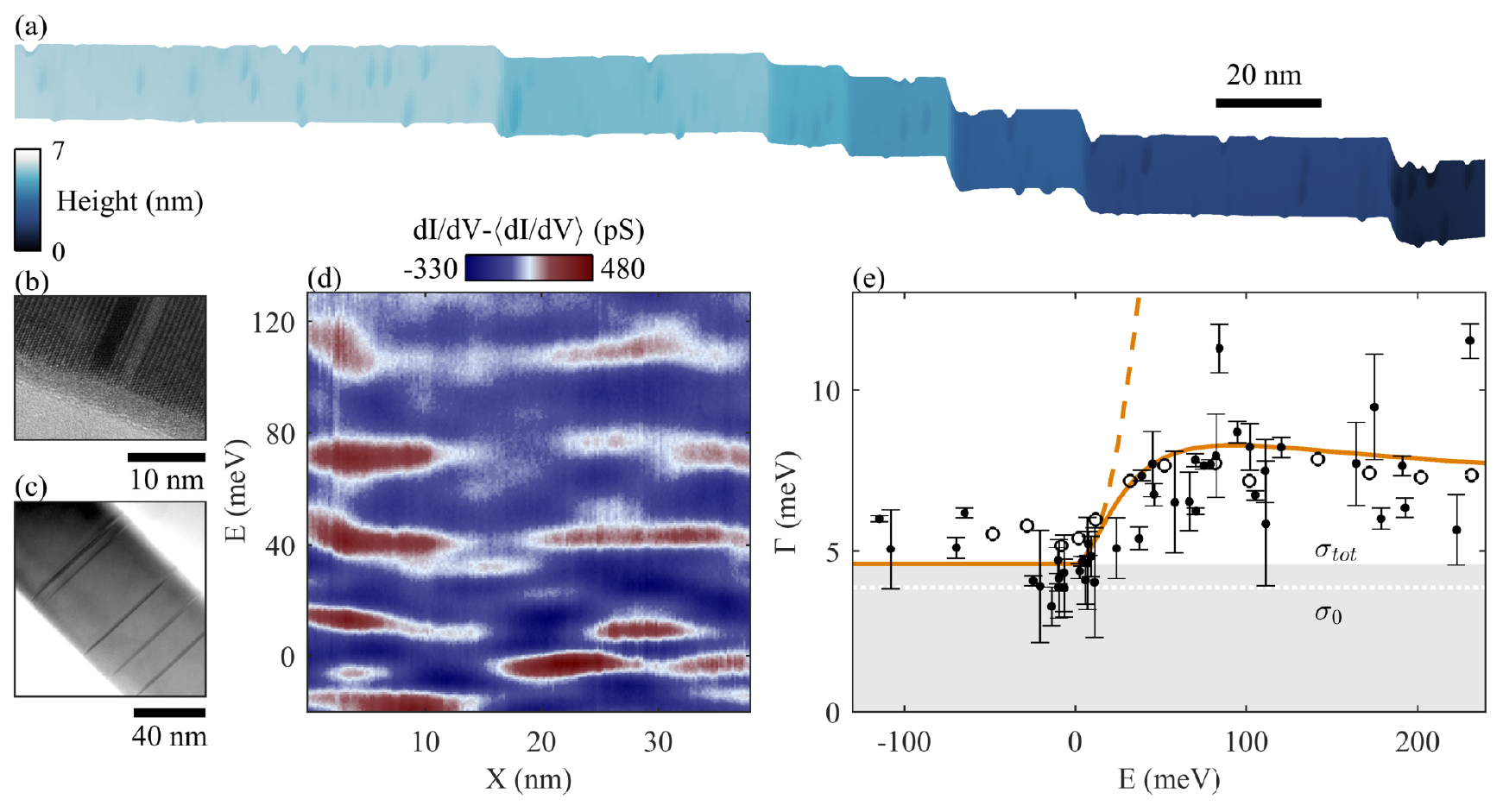} 
\caption{\label{Fig 3} Non-monotonic phase coherence time  extracted from Fabry-P\`{e}rot resonators. (a) Topographic image showing a series of adjacent stacking faults on the top facet of the nanowire. (b) High resolution TEM micrograph demonstrating that a single stacking fault induces atomic displacement. (c) TEM micrograph of a nanowire from the same growth batch showing a similar distribution of stacking faults. (d) Typical dI/dV line-cut along a terrace terminated by adjacent stacking faults (mean spectrum subtracted) showing quantized resonances. Up to 20\% of the mean LDOS is modulated. (e) Energy broadening of the quantized resonances $\Gamma\left(E\right)$ as a function of their energy (black dots). The error bars originate in a two-stage fitting procedure described in Appendix \ref{E2}. At the Fermi energy the broadening $\sigma_{tot}$ is attributed to instrumental contributions of temperature and probing AC amplitude, $\sigma_{0} = 3.9$ meV (up to dotted line) and the transmissivity of the stacking faults, $\sim 1.1$ meV, setting a lower bound on the reflection coefficient of $\left|R\right|^{2}\geq 0.8 \pm 0.1$. The excess energy dependent broadening (above $\sigma_{tot}$) is attributed to the finite coherence time of the electrons $\tau_{\varphi}$. Both the model (solid line) and the coherence length data from Fig. \ref{Fig 2}c (circles) agree on the non-monotonic evolution of the electronic phase coherence and deviate from the low-energy evolution (dashed line). }
\end{figure}

A distinct imprint of the non-monotonic relaxation profile was measured concurrently away from the nanowire end by studying Fabry-P\`{e}rot like structures formed by adjacent crystallographic stacking faults. Topographically, such a series of adjacent stacking faults appears as consecutive parallel corrugations on the nanowire surface, whose heights do not register with the size of the unit cell (Fig. \ref{Fig 3}a). Both the structure of the stacking faults and their lateral distribution visualized in STM agree with those seen in transmission electron microscopy of nanowires harvested from the same growth (Fig. \ref{Fig 3}b-c). The dI/dV map taken in between the stacking faults displays faint resonances (Fig. \ref{Fig 3}d) with a fairly symmetric spatial structure on top of the continuous background, as expected for electrons in a leaky resonator (see Appendix \ref{E3}).  We characterize these resonances by plotting their energy width $\Gamma$ versus their peak energy (Fig. \ref{Fig 3}e, more details can be found in Appendix \ref{E2}). To extract the energy dependence of the relaxation rate we account for the various contributions to the resonance width; Electrons injected at the Fermi energy do not relax. The width of the resonance there $\Gamma\left(E_{F}\right)$, is contributed by the finite time the electrons reside inside the resonator due to the leakiness of its boundaries, as well as by instrumental contributions due to thermal broadening ($\sim 1$ meV) and finite probing amplitude (3 meV). By subtracting out the instrumental contributions and assuming an energy independent scattering barrier \cite{Burgi1999,Seo2010} we find a high reflection coefficient of $\left|R\right|^{2}\geq 0.8 \pm 0.1$ for scattering of an electron off a stacking fault (more details in Appendix \ref{E3}). 

The remaining energy-dependent contribution to the resonance broadening, $\widetilde{\Gamma} \left(E\right) = \Gamma\left(E\right) - \Gamma\left(E_{F}\right)$, is attributed to the finite life-time  of the injected hot electrons before relaxing towards the Fermi sea. The extracted relaxation rate, $\tau\left(E\right) \sim \hslash/2\Gamma\left(E\right)$, is increasing above $E_{F}$ due to the increase in phase space of states to interact with (Fig. \ref{Fig 3}e, dashed line).  Remarkably, instead of continuing its monotonic increase with energy, the relaxation rate saturates and starts decreasing (solid line). The hot electrons velocity $v$ extracted from the measured dispersion curve (Fig. \ref{Fig 1}e), relates the observed non-monotonicity of the relaxation rate $\left(1/\tau_{\varphi}\right)$ to the similar trend we detect in electrons coherence length (Fig. \ref{Fig 2}c, \ref{Fig 3}e) via $L_{\varphi} \sim v \tau_{\varphi}$. This ballistic formalism can be used because the end of the nanowire is free from impurities (see Fig. \ref{Fig 1}c) \cite{Burgi1999}. Furthermore, surface defects, imaged at remote regions, have significantly smaller scattering amplitude than the one exhibited by the physical end of the nanowire and do not seem to embed a superimposed QPI pattern. Both mechanisms, that reflect relaxation induced phase decoherence of hot-electrons in different manners, agree on the non-monotonic behavior, and find a similar energy scale of 80 meV, at which the relaxation rate is maximal. Discrepancy in the overall magnitude between the two is attributed to a slight variation in the suspension of the nanowires over the Au substrate, discussed below. We thus firmly conclude that 1D hot electrons in semiconducting nanowires indeed exhibit an energy regime in which they regain phase coherence with increasing energy.

\section{Theoretical model \label{Theory}}

We now turn to discuss the physical origins behind the extended phase coherence of the lowest subband, and then its detected non-monotonic behavior as a function of energy. For more details see model derivation in Appendix \ref{D}. For an injected hot electron to embed an interference pattern in the LDOS it must not relax throughout its scattering path. In 1D nanowires hot electrons lose their phase coherence predominantly by interacting with the bath of cold electrons that occupy the Fermi-sea. This electron-electron interaction promotes relaxation of the system towards its ground-state. We concentrate on relaxation facilitated by the Coulomb interaction since in 1D excitations of optical phonons carry negligible momentum and acoustic phonons are too slow to effectively interact with the electrons \cite{Ji-YongPark2004} (thermally excited phonons are practically absent at 4.2 K).

\begin{figure}
\includegraphics{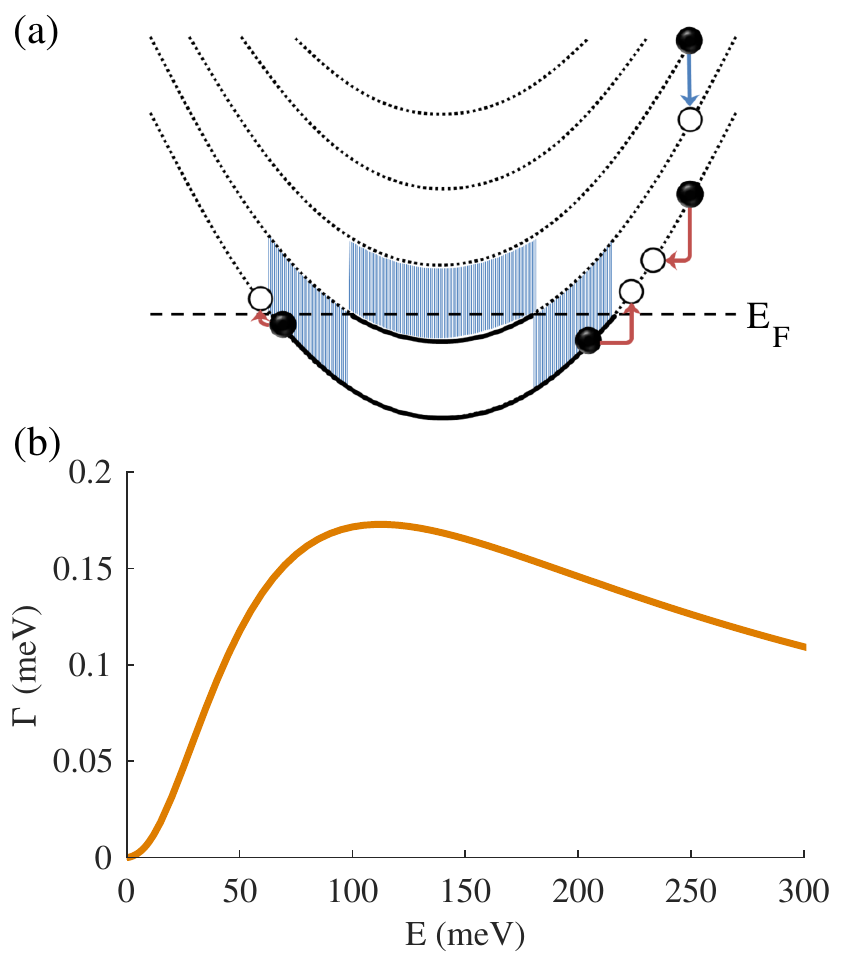} 
\caption{\label{Fig 4} Modelling the relaxation rate of a 1D electronic subband. (a) Schematic comparison of the slow 3-body relaxation of hot-electrons on the lowest subband, that excites both co- and counter-propagating electron-hole pairs, and the fast 2-body relaxation of hot-electrons on higher subbands (red and blue arrows, respectively). (b) The calculated relaxation rate of a single subband with quadratic dispersion. The dispersion was chosen to have the same Fermi momentum and Fermi velocity as the nanowire'’s lowest subband.  }
\end{figure}

We find that the interaction induced relaxation processes available for hot electrons in the lowest subband are markedly different from those available at higher subbands. Hot electrons occupying higher subbands decay predominantly via inter-band relaxation processes (blue arrow in Fig. \ref{Fig 4}a). These involve excitation of a single electron-hole pair off the Fermi-sea with essentially no momentum transferred between the two, namely $q=0$. Therefore, the phase space available for such processes is relatively broad (shaded region). We calculate an upper bound for their resulting phase coherence length of $L_{\varphi} \leq 10$   nm (see Appendix \ref{D1}). In contrast, hot-electrons occupying the lowest subband relax only via intra-band 3-body scattering processes, much like hot electrons in a single-subband nanowire. These 3-body processes necessitate exchanging both energy and momentum with the Fermi-sea (red arrows in Fig. \ref{Fig 4}a), which for any finite curvature in the dispersion involves excitation of both co- and counter-propagating electrons \cite{Imambekov2012,Lunde2007,Karzig2010a,Ristivojevic2013}. The phase-space for such 3-body relaxation processes is much more restricted and the strength of the high order interaction term is significantly weaker, rendering the relaxation within the lowest subband substantially slower. This clarifies why the imaged intensity of this subband dominates all other subbands in Fig. \ref{Fig 1}e. 

It is this 3-body relaxation process of 1D hot electrons that exhibits the non-monotonic energy evolution. The relaxation rate calculated via Fermi'’s golden rule is contributed both by the available phase-space of states as well as by the strength of the interaction. We find that the available phase space for effective interaction is limited by a finite cut-off in the momentum transferred in that scattering process $ q \sim 1/d $. Above the energy corresponding to this characteristic momentum the interaction among the 1D electrons $V(q)$ is rapidly suppressed:

\begin{equation*}
V\left( q \right) \sim \left\{ {\begin{array}{*{20}{c}}
  {\ln \left( {qd} \right)\,\,,\,\,q \ll {1 \mathord{\left/
 {\vphantom {1 d}} \right.
 \kern-\nulldelimiterspace} d}} \\ 
  {{{\left( {qd} \right)}^{ - 2}}\,\,,\,\,q \gg {1 \mathord{\left/
 {\vphantom {1 d}} \right.
 \kern-\nulldelimiterspace} d}} 
\end{array}} \right.
\end{equation*}

Consequently, above a certain energy threshold the hot electron ceases to interact with deeper states in the Fermi sea because this involves an exceedingly large momentum transfer, $ q >> 1/d$. The increase in the relaxation rate saturates accordingly. Remarkably, the relaxation rate starts \textit{decreasing} beyond this saturation level. The reason is the increasing velocity mismatch between that of the injected hot-electron $v_{i}$ and the Fermi-velocity, $v_{F} \sim 1 \times 10^{6}$ m/sec, which characterizes the electron-hole pairs excited off the Fermi-sea in the 3-body relaxation process. This suppression can be interpreted as resulting from the limited time window the co-propagating scattered electrons with different velocities have for interaction. Thus, the combination of saturated phase-space growth with decreasing interaction leads to the observed non-monotonic relaxation profile in 1D, as demonstrated in Fig. \ref{Fig 4}b.

\section{Discussion \label{Discussion}}

The model captures the behavior we find in our experiments remarkably well –- both the energy at which it assumes its maximal value at about 80 meV above $E_{F}$ as well as the order of magnitude of the relaxation rate. Note, that in the fits shown in Figs. \ref{Fig 2}c and \ref{Fig 3}e we have also accounted for inter-subband scattering processes in which the hot-electron on the lowest subband excites electron-hole pairs at higher occupied subbands (exchange processes are still negligible as they necessarily involve large momentum transfers  ). Intriguingly, we find that as the Fermi-sea becomes shallower, its finite depth may saturate the growing phase space even before the 1D Coulomb interaction does so (Appendix \ref{D3}). All parameters except for the fitted overall prefactor are extracted from the experiment. Considering a more complex cross-sectional wave function distribution, for example to account for a surface charge accumulation, may somewhat shift the transition energy and the magnitude of the effect, but will not change the qualitative phenomenon. The prefactor is highly sensitive to the ratio between the wave-function'’s lateral extent and the dielectric constant. When substituting the nanowire'’s physical diameter the fit yields a dielectric constant of order unity. It is thus much closer to that of vacuum than to that of bulk InAs, $\epsilon_{bulk} \sim 15$, asserting that the Coulomb interaction is hardly screened. This is consistent with the electronic wavefunction being localized closer to the nanowire circumference and the nanowires being suspended over the rough Au substrate. Accordingly, the discrepancy in the overall magnitude of the two experimental methods studied, translates to a slight fluctuation in the extracted effective dielectric ($\epsilon_{end} \sim 1$ versus $\epsilon_{center} \sim 2)$ that is attributed to small variations in the suspension between the nanowire end and center. These values further signify that 1D electrons in suspended InAs nanowires are moderately interacting with a Coulomb to kinetic energy ratio of $r_{s} \sim 2-3$.  

Intriguingly, our model predicts that in 1D materials with extended quadratic dispersion, such as boron-nitride nanotubes \cite{Blase1994}, or simply for free electrons in 1D the relaxation rate will continue to decrease and become vanishingly small with increasing energy as the velocity mismatch continues to grow. Furthermore, the more nonlinear the dispersion is, such as in silicon and GaAs, the faster the relaxation rate will decrease with energy. On the other hand, for any electronic band of finite width the recovery of the phase coherence will eventually saturate at the inflection point of the dispersion. Further above this energy, once the velocity of the ultra-hot electrons drops below $v_{F}$ we find that the phase coherence length will start decreasing since fast two-body relaxation processes become accessible. Nevertheless, the magnitude of this decrease will be suppressed since these two-body relaxation processes necessarily involve a finite momentum transfer that again increases with increasing energy thus suppressing the Coulomb interaction and their contribution to the relaxation rate.

The discovered energy window of extended phase coherence and slow relaxation lends new opportunities in coherent manipulations and applications of ultra-hot 1D electrons in semiconducting nanowires that were otherwise thought to be highly unstable. A concrete example is given by photovoltaic quantum dots embedded in semiconducting nanowires whose performance is limited by the energy loss due to relaxation of the photo-electrons as they are transported to the leads \cite{Beard2014}. On a more general ground, our finding of the new regime of extended phase coherence in 1D can stabilize quasi-particles in 1D, sprouting novel theoretical descriptions in the spirit of Landau'’s paradigm for interacting ultra-hot electrons, and harness their extended coherence for various quantum applications and technologies.

\begin{acknowledgments}
HB acknowledges funding by the Israeli Science Foundation (ISF), the Minerva Foundation, and the European Research Council (ERC, project ‘TOPONW’). HS acknowledges partial funding by Israeli Science Foundation (grant No. 532/12 and grant No. 3-6799), BSF grant No. 2014098 and IMOS-Tashtiot grant No. 0321-4801. We are grateful to Maria-Theresa Rieder, Yuval Oreg, Felix von Oppen and Ady Stern for many fruitful discussions. We thank Hadar Eizenshtat for his contribution to the measurements. This work was performed in part at the Aspen Center for Physics, which is supported by National Science Foundation grant PHY-1066293.
\end{acknowledgments}

\appendix

\section{Materials and methods}

\subsection{\label{A1} Molecular beam epitaxy growth of InAs nanowires}

\begin{figure}
\includegraphics{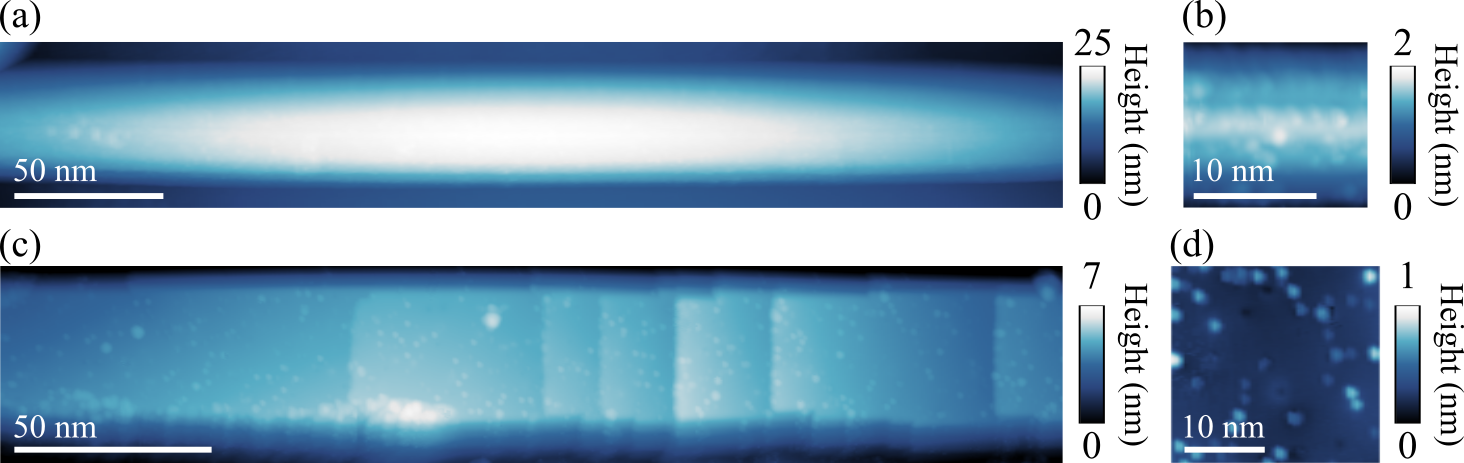} 
\caption{\label{Fig 5} (a) STM topography of a nanowire grown without faceting exhibiting round cross-section.  (b) Zoomed-in topography of (a). The round cross-section in conjugation with the lattice discretization yields a corrugated stepped surface filled with randomly oriented dangling bonds. (c) A large-scale topography of a wide facet on top of a wire with multiple stacking faults (same as in Fig. \ref{Fig 3}). (d) Zoomed-in topography of one of the terraces. The average distance between impurities is couple of nm. }
\end{figure}

InAs nanowires were grown on (111)B InAs substrate by the gold assisted vapor liquid solid (VLS) technique in a high purity molecular beam epitaxy (MBE) growth system. The initial gold layer is evaporated in-situ after removal of the oxide layer. In order to facilitate a clean harvesting (see below) we grow the nanowires at relatively low density. This is obtained by nucleation $70^{\circ}$ C above the growth temperature which is then reached by ramping down at $5^{\circ}$ C per minute.  Under such conditions the normally clean Wurtzite nanowires grow with occasional stacking faults along the growth axis (Figs. \ref{Fig 3}b-c, \ref{Fig 5}). As mentioned in Sec. \ref{Experiment_A} flat atomic facets are essential for STM spectroscopy. In order to facilitate the formation of facets around the normally round nanowires, after axial growth of one hour at $ \sim 500^{\circ}$ C the temperature was ramped down by $ \sim 70^{\circ}$ C in order to enhance side growth for 15 additional minutes. This produced relatively thick InAs nanowires with prominent faceting. The significance of the faceting, along with an impression of the typical distribution of surface impurities found on the clean facets (after transfer in UHV), is demonstrated in Fig. \ref{Fig 5}. 

\begin{figure}
\includegraphics{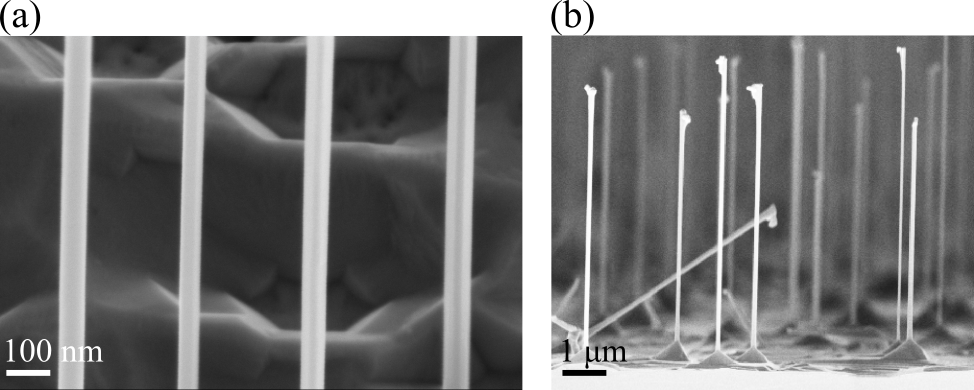} 
\caption{\label{Fig 6} Scanning electron microscopy images of InAs nanowires. (a) The faint shadows along the nanowires axis are atomically flat facets. (b) nanowires on their (111)B InAs substrate. The tilted nanowire has randomly grown in the (100) direction.}
\end{figure}

The imaged side corrugations of the nanowires, seen for example on Fig. \ref{Fig 1}a, result from the convolution of the profile of the nanowire with that of the STM tip, and therefore do not reflect their true diameter. The partial suspension does not allow its extraction from the height difference from the top surface to the substrate. Scanning electron microscopy (SEM) images of the same batch (Fig. \ref{Fig 6}) find a rather uniform diameter, $d=70\pm10$ nm.  
.

\subsection{\label{A2} Au substrates}

\begin{figure}
\includegraphics{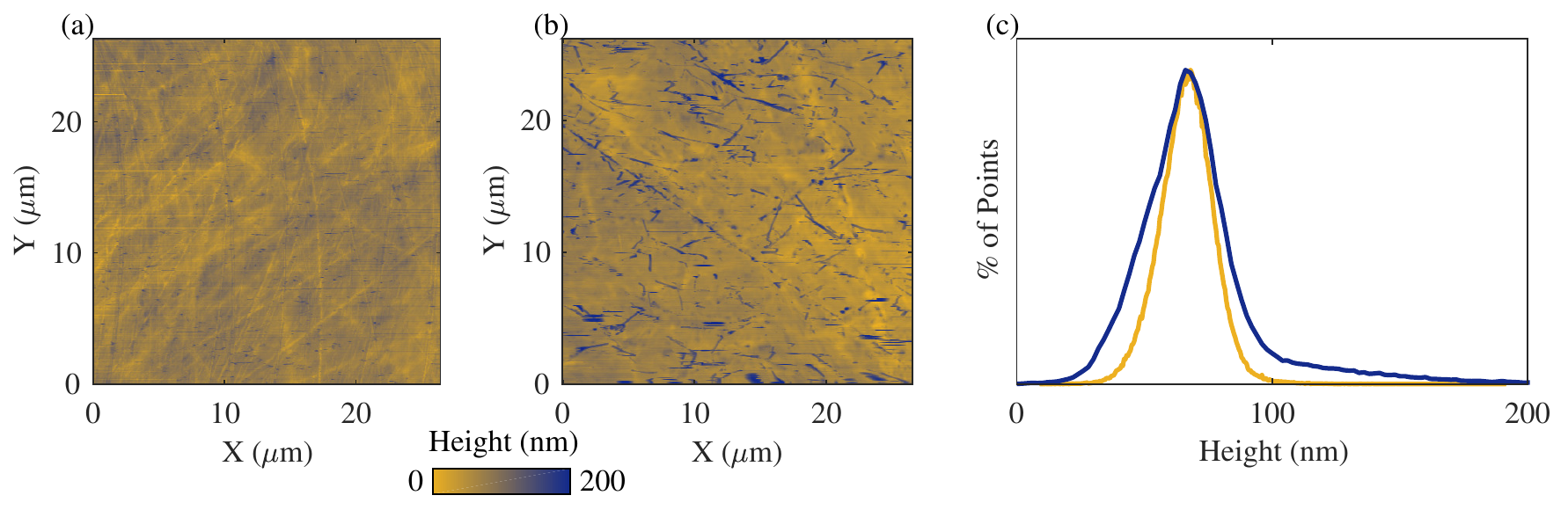} 
\caption{\label{Fig 7} AFM topographies and statistical analysis of surface roughness Topography of  clean (a)  and nanowire covered (b) Au surfaces. (c)  Height distribution of the topography in (a) (orange) and (b) (purple). The roughness can be characterized by the variance of these distributions. }
\end{figure}

Preparation of the Au single crystals for nanowire deposition includes an ex-situ polishing and in-situ sputtering and annealing procedures. We first embed the Au single crystals in crystal bond to facilitate their mechanical polishing in alumina suspension with decreasing grain size (down to 50 nm). The crystal bond is then removed from the single crystal by sonicating it in warm solvent. We characterized the polished surface roughness in an atomic force microscope (AFM) before (Fig. \ref{Fig 7}a) and after (Fig. \ref{Fig 7}b) nanowire deposition. Although it is shiny and reflective to the naked eye, the resulting polished Au surface is somewhat scratched on a sub-micron scale. The surface roughness before nanowire deposition (orange line in Fig. \ref{Fig 7}c) is on the order of 40 nm, comparable to the mean nanowire diameter. Prior to nanowire deposition the polished Au substrates are further Ar-sputtered and annealed under UHV conditions. This procedure produces contaminant-free crystalline Au surfaces but does not heal the surface roughness caused by the polishing. InAs nanowires are subsequently deposited on them by mechanically pressing the clean Au surface against the nanowire growth substrate. This procedure seems to increase the surface roughness, as illustrated by the wider height distribution in Fig. \ref{Fig 7}c (purple line). The distribution is broadened to $\sim 60$ nm and an asymmetric tail appears due to the presence of nanowires.  This roughness agrees with the height variation we measure in topography mode in STM. 

\subsection{\label{A3} Ultra-high vacuum suitcase}

\begin{figure}
\includegraphics{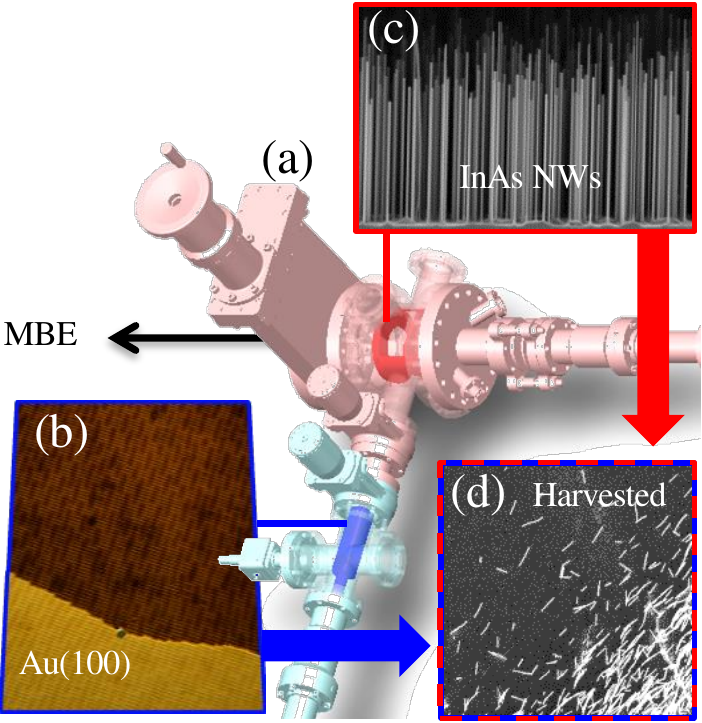} 
\caption{\label{Fig 8} UHV growth, harvest and transfer of nanowires. (a) UHV suitcase (blue) and harvesting chamber (red) in which we disperse the grown nanowires on a Au surface and transfer them under UHV from the MBE to STM chamber. (b) Topography of freshly prepared Au(100) substrate. (c) SEM image of vertically growing InAs nanowires. (d) SEM image of harvested nanowires dispersed over the Au substrate.}
\end{figure}

The nanowires were transferred from the MBE growth chamber to a Unisoku low-temperature STM under ultra-high vacuum (UHV) conditions ($ \sim 10^{-10}$ Torr), in a designated portable suitcase. To achieve this vacuum level the suitcase was equipped both with an ion pump and a non-evaporable getter pump. Nanowire harvesting is performed in the suitcase by gently pressing freshly prepared Au crystals against the nanowires growth substrate. In this process some of the nanowires break (Fig. \ref{Fig 8}) and attach to the Au surface, presumably by van der Waals forces. The whole transferring procedure lasts less than an hour.

\subsection{\label{A4} STM measurements}
 
\begin{figure}
\includegraphics{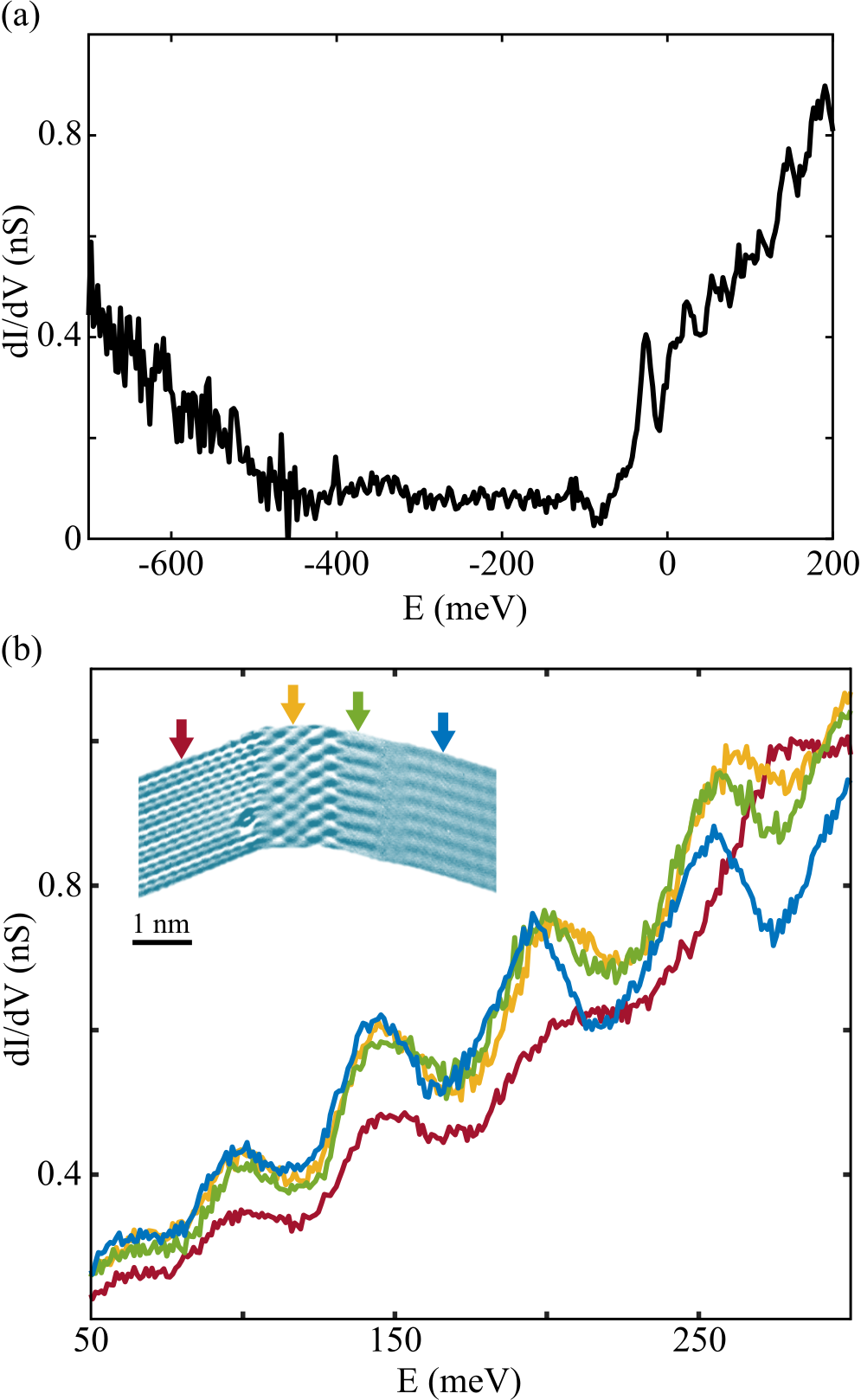} 
\caption{\label{Fig 9} (a) Electronic spectrum of InAs nanowire. $E_F$ is $\sim 75$ meV above the conduction gap minimum, and the semiconducting gap is about 350 meV. Resonances indicating the 1D sub-bands are seen in both the conduction and valance bands. (b) Van-Hove singularities from spectra taken on different crystal termination. locations across the nanowire are shown in inset by arrows of corresponding color over the topography.}
\end{figure}

 We scan the surface with a Pt-Ir tip in search of the deposited nanowires. The tip can be reconditioned and tested on the clean Au terraces to ensure stability, metallic behavior and reproducible results. dI/dV spectra were obtained using standard lock-in measurement. Typical parameters are Lock-in frequency of 733 Hz, AC amplitude of 3 mV, parking bias of 250 meV, and current set-point of 250 pA.
Fig. \ref{Fig 9}a shows a broader point spectrum of a nanowire. This spectrum shows the full semiconducting gap of InAs bounded by the onset of the valence and conduction bands. This spectrum is indifferent to the particular crystal termination on which it was measured, as demonstrated in Fig \ref{Fig 9}b. The stability of the tunneling junction deteriorated at higher biases which limited our energy window to about $\pm 0.5$ eV, which is quite considerable.

A formation of quantum dot states, caused by tip-induced band bending, was previously reported in STM studies of bulk InAs \cite{Dombrowski1999}. We rule out this possibility in our setup since that would have generated a quantized QPI mode and relies on $E_{F}$ lying a few meV's from the bottom of the band, whereas we find a continuous QPI mode whose minimum lies $\sim 75$ meV below $E_{F}$ (Fig. \ref{Fig 1}e). A fundamental difference in our setup is the presence of the Au substrate that can efficiently screen the invasiveness of the macroscopic tip. 

\section{\label{B} Ab-initio calculation of subband quantization in Wurtzite InAs nanowires}

To calculate the electronic structure of bulk Wurzite InAs, we have employed density-functional theory (DFT) calculations, which is implemented in the Vienna Ab-initio Simulation Package (VASP) \cite{Kresse1996}. The hybrid-functional (HSE06) \cite{Heyd2003} was adopted for the exchange-correlation effect and the spin-orbit coupling was included. We extracted tight-binding parameters from the DFT bulk calculations by projecting the Bloch wave functions to localized Wannier functions \cite{Marzari1997} that correspond to the In-s and As-p atomic-like orbitals. The Hamiltonians of nanowires were constructed based on these Wannier parameters.

\begin{figure}
\includegraphics{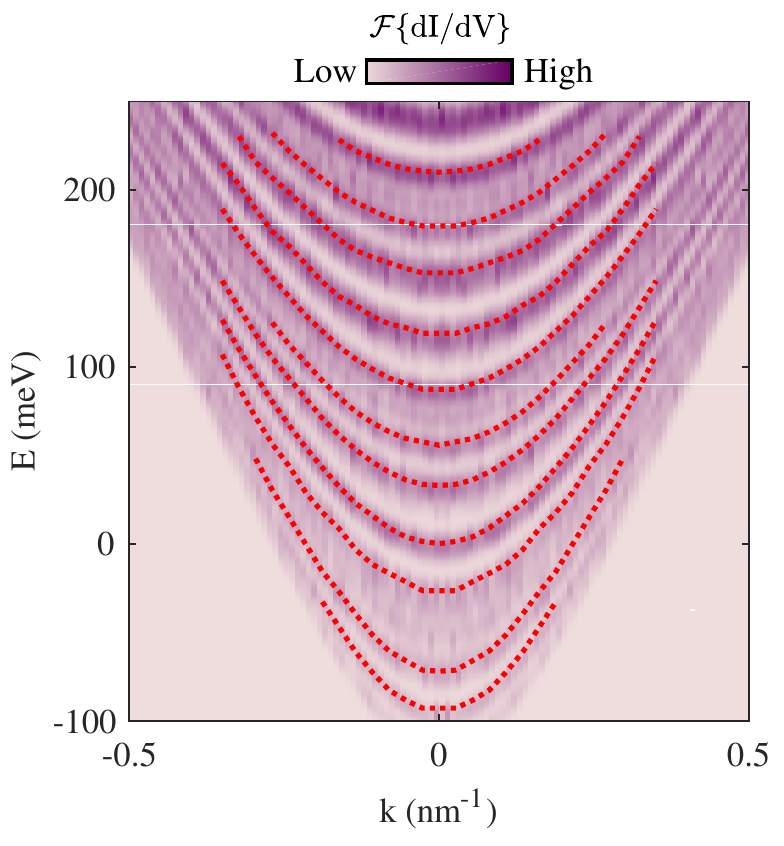} 
\caption{\label{Fig 10} Tight binding calculation of InAs nanowires band structure. Dotted lines mark some of the calculated bands. }
\end{figure}

We determine the bandstructure of the system theoretically by means of a kernel polynomial approximation.\cite{Weisse2006} Using the tight-binding Hamiltonian obtained from ab-initio calculations, we define a nanowire geometry which is translationally invariant along the $c$-axis and has a hexagonal cross-section. For the plots of Fig. \ref{Fig 1}b and Fig. \ref{Fig 10} we have used a nanowire with an outer diameter $d\simeq 72.8\, {\rm nm}$. This corresponds to a momentum-dependent Hamiltonian $H(k_z)$ with 21931 sites, each of which describes the $X$-orbital degrees of freedom of the two atomic species (In and As), as well as spin.

The large size of the Hamiltonian matrix makes direct diagonalization impractical. Instead, we approximate the bandstructure by determining the momentum resolved density of states, $\rho(k_z, E)$, using a kernel polynomial method. For each value of $k_z$, the density of states can be written in terms of the eigenvalues of the Hamiltonian, $E_n(k_z)$, as

\begin{equation}\label{eq:dos}
 \rho(k_z, E) = \sum_{n} \delta(E - E_n(k_z)).
\end{equation}

Rather than determining the eigenvalues directly, we expand \eqref{eq:dos} in a series of Chebyshev polynomials

\begin{equation}\label{eq:cheb}
 T_n(x) = \cos(n\, {\rm arccos}(x)).
\end{equation}

The polynomials \eqref{eq:cheb} obey the recursion relations

\begin{equation}
 T_{n+1}(x) = 2x \, T_n(x) - T_{n-1}(x),
\end{equation}
with $T_0(x)=1$ and $T_1(x)=x$, and are defined in the interval $x\in[-1,1]$. Due to this latter constraint, the kernel polynomial method requires rescaling the Hamiltonian such that its spectrum is contained in the interval $[-1,1]$:

\begin{equation}
\widetilde{H} = \frac{H - b}{a},
\end{equation}
with energies
\begin{equation}
\widetilde{E} = \frac{E - b}{a}.
\end{equation}

We set $a=6.35$ eV, $b=3.11$ eV, and expand the rescaled density of states
\begin{equation}\label{eq:dosrescaled}
 \widetilde{\rho}(k_z, E) = \sum_{n} \delta(E - \widetilde{E}_n(k_z))
\end{equation}
in an infinite series
\begin{equation}\label{eq:expansion}
 \widetilde{\rho}(k_z, E) = \frac{1}{\pi\sqrt{1-E^2}}\left[ \mu_0 +2\sum_{n=1}^\infty \mu_n T_n(E) \right],
\end{equation}
with the expansion coefficients
\begin{equation}\label{eq:coeff}
 \mu_n = {\rm Tr}\, T_n \left[ \widetilde{H}(k_z) \right],
\end{equation}
where ${\rm Tr}$ denotes the trace.

The infinite sum appearing in \eqref{eq:coeff} is truncated by keeping only the first $N$ terms
\begin{equation}\label{eq:truncation}
 \widetilde{\rho}(k_z, E) \simeq \frac{1}{\pi\sqrt{1-E^2}}\left[ \mu_0 +2\sum_{n=1}^{N-1} \mu_n T_n(E) \right],
\end{equation}
an approximation which leads to fluctuations in the density of states, also known as Gibbs oscillations. The latter can be reduced by modifying the expansion coefficients as $\mu_n\to g_n\mu_n$, where we make the choice
\begin{equation}
 g_n = \frac{(N-n+1)\cos\displaystyle{\frac{\pi n}{N+1}}+\sin{\frac{\pi n}{N+1}}\cot{\frac{\pi}{N+1}}}{N+1},
\end{equation}
corresponding to the so-called Jackson kernel.\cite{Weisse2006}

The numerical results of Figs.\ref{Fig 1}b and \ref{Fig 10} are obtained using $N=8192$ polynomials for each value of $k_z$, and the coefficients \eqref{eq:coeff} are determined using a stochastic evaluation of the trace
\begin{equation}\label{eq:coeffrnd}
 \mu_n = {\rm Tr}\, T_n \left[ \widetilde{H}(k_z) \right] \simeq \frac{1}{R}\sum_{r=0}^{R-1} \langle r | T_n(\widetilde{H}(k_z)) | r \rangle.
\end{equation}
Here, $|r\rangle$ are random vectors with entries drawn from the Gaussian distribution with zero mean and unit variance. Each coefficient was computed using $R=10$ random vectors.

\section{\label{C} Visualizing subbands via scattering off point impurities}

\begin{figure}
\includegraphics{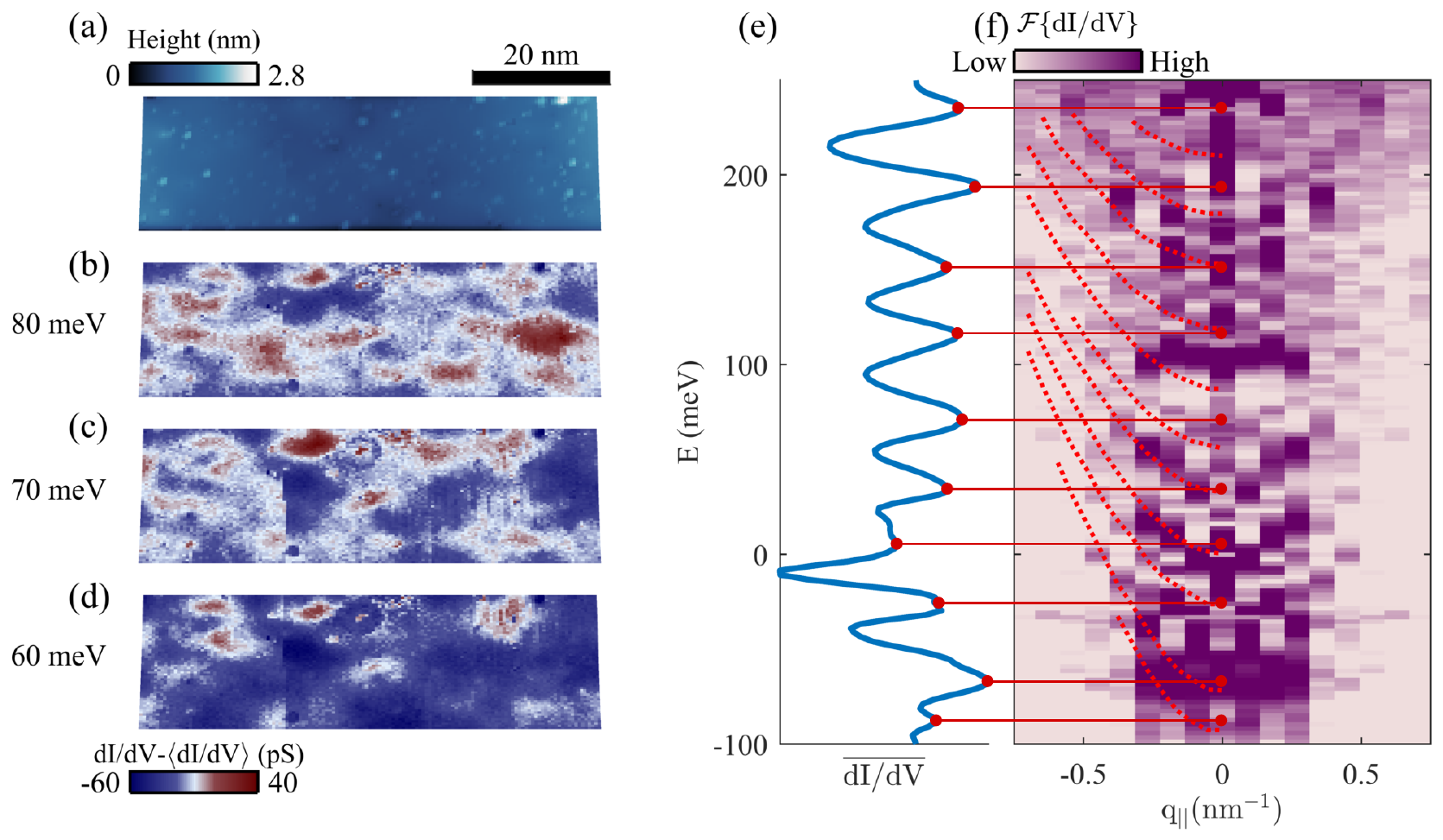} 
\caption{\label{Fig 11} Quasi-particle interference imaging of the quantized bands in InAs nanowires (a) Topography of the nanowire surface with various surface impurities that serve as scattering centers. The horizontal direction is parallel to the direction of the nanowire. (b)-(d), Differential conductance (dI/dV) maps at 80, 70 and 60 meV showing dispersive QPI patterns manifested in spatial fluctuation around the mean value at each energy. Comparing the spatial standard deviation of the LDOS to the mean spectrum yields that roughly 5\% of the LDOS is modulated.  (e) Mean spectrum showing Van-Hove singularities. The spectrum was obtained by spatially averaging the dI/dV maps and subtracting a monotonic smooth background for visibility. (f) Fourier transform of the dI/dV maps with respect to the direction of the nanowire. The red dotted curves are copied from  Fig. \ref{Fig 10} (the x-axis was scaled to represent momentum transfer $q=2k$), and demonstrate the close resemblance between the calculated and the measured band structure.  The solid red lines indicate that the bottom of each band originates from a specific Van-Hove singularity.}
\end{figure}

We visualized the 1D electronic structure by measuring the QPI patterns embedded in the LDOS by the scattered electrons. The scatterers in this segment of the nanowire are point-impurities adsorbed to the surface during the growth or the sample transfer. Spectroscopic maps were measured over the upper facet of the segment of a nanowire, whose topography is shown in Fig. \ref{Fig 11}a. The imaged spatial fluctuations of the LDOS about the mean value (Fig. \ref{Fig 11}b-d) are clearly energy dependent, signifying that these fluctuations result from the electronic scattering off point-impurities. The spatially averaged dI/dV value of these maps shows the expected spectrum quantization into a series of Van-Hove singularities (Fig. \ref{Fig 11}e). The momentum transferred between the impinging and scattered states, $q = k_{i} -k_{f}$, is embedded in the spatial fluctuations. By performing a Fourier transform along the nanowire axis at the different energies we capture a series of dispersing bands shown in Fig. \ref{Fig 11}f. To the best of our knowledge, This is the first time 1D channels were visualized in semiconducting nanowires. The measured QPI agree remarkably well with the ab-initio calculation of the spectrum (dotted lines). Fitting a quadratic term at the bottom of each band yields an effective mass ($ m^{*} = 0.05 \pm 0.01\mathrm{m_{e}}$) that resembles that of bulk InAs ($m^{*} \sim 0.02 \mathrm{m_{e}}$). The peak of the Van-Hove singularities seen in the average spectrum coincide with the minima of the visualized sub-bands (marked by solid red lines), serving as an important consistency check.

\section{\label{D} Modeling hot-electron relaxation}

\subsection{\label{D1} Two-body inter-subband relaxation}

To estimate the strength of the faster two-body processes that allow hot particles in higher subbands to relax via inter-band transitions, we use the first order Golden rule expression 

\begin{equation}
\Gamma_{\rm int}=\sum_{k_{2}q}\sum_{\sigma_{2}\sigma_{1'}\sigma_{2'}}\frac{2\pi}{\hbar}|\langle 1'2'|V|12\rangle|^2 n_{2}(1-n_{1'})(1-n_{2'})\delta(\varepsilon_\mathrm{i}-\varepsilon_\mathrm{f})\,,\label{eq:out_inter}
\end{equation}

Where $i=1,2$ labels the states and denotes initial states at momenta $k_{i}$ and spin $\sigma_{i}$, and $j=1',2'$ denotes final states at momenta $k_{j'} = k_{j} + q_{j}$ and spin $\sigma_{j'}$. The Fermi-Dirac distribution functions $n_{i}$  describe the condition of occupied initial and empty final states and are evaluated at zero temperature in the following. The Coulomb amplitude takes the form $\langle 1'2'|V|12\rangle=\left(V^\mathrm{int}_q\delta_{\sigma_1,\sigma_{1'}}\delta_{\sigma_2,\sigma_{2'}}-V^\mathrm{int}_{k_{1'}-k_2}\delta_{\sigma_1,\sigma_{2'}}\delta_{\sigma_2,\sigma_{1'}}\right)/L$, where $V^\mathrm{int}_q$ is the Coulomb matrix element for scattering between two subbands while transferring momentum $q$. We assume that the two subbands involved have the same dispersion which is only shifted by an energy offset, which is well supported by the ab-initio calculation. Moreover, we assume that the lower subband is partially filled for $|k_2|<k_F$ while the upper band is entirely empty. Note that the latter condition can effectively be reached by choosing the band of the excited particle with momentum $k_2$ as the highest occupied band at that momentum. Thereby the phase space of $k_2$ extends from $-k_F$ to $k_F$  (see Fig. \ref{Fig 4}a for a system with two occupied bands).   
After summing over the spins and resolving the energy delta function (which yields the condition $q=0$) we obtain
\begin{equation}
\Gamma_{\rm int}=\frac{2}{\hbar}\int_{-k_F}^{k_F} \frac{d k_2}{2\pi}\frac{1}{|v(k_1)-v(k_2)|}\left([V^\mathrm{int}_0]^2+[V^\mathrm{int}_{k_1-k_2}]^2-V^\mathrm{int}_0V^\mathrm{int}_{k_1-k_2}\right)\,,
\end{equation}

Since an inter-band scattering involves transverse momentum transfers $\sim 1/a$ ($a$ being the wire radius), we estimate $V_0^\mathrm{int}\sim V_{1/a}$. For large injected momenta $k_1-k_F\gg 1/a$,  the direct terms of the interaction $V^\mathrm{int}_0$ will dominate. Assuming a quadratic dispersion for the occupied part of the band, i.e. $v(k_2)=k_2/m$, we can then perform the exact momentum integration which yields

\begin{equation}
\Gamma_{\rm int}=\frac{k_F}{\hbar v_F\pi}\log\left(\frac{v(k_1)+v_F}{v(k_1)-v_F}\right)[V^\mathrm{int}_0]^2\,,
\end{equation}

When using the parameters obtained from a fit of $\varepsilon\left(k\right) = \hslash v \sqrt[]{k^{2}+k_{0}^{2}} + \varepsilon_{0}$  to the measured dispersion ($\varepsilon\left(k_{F}\right) = 0$), yielding $v = 1.4 \times 10^{6}$ m/sec, $k_{0} = 0.19 \mathrm{nm^{-1}}$, $\varepsilon_{0} = -250$ meV,  we finds very fast rates of the order of electron volts. Indeed, such high rates might signal a breakdown of the perturbation theory in that limit. For $V^\mathrm{int}_0\lesssim v_F$, one should be well inside the perturbative limit. Since in that case $\Gamma_\mathrm{int}\sim \varepsilon_F/\hbar$, it seems reasonable that the typical time scales for interband relaxation are at least of the order of the Fermi energy. For $\varepsilon_{F} \sim 75$ meV, the upper limit on the scattering time is $\sim 10$ ps, which leads to $L_{\varphi} \leq 10$ nm.

\subsection{\label{D2} Three-body intra-subband relaxation}

In clean one dimensional electron systems energy and momentum conservation strongly restricts possible relaxation mechanisms. In particular, when taking into account the finite curvature of the electron dispersion simple two-body relaxation processes (as in standard two or three dimensional Fermi-liquid theory) cannot lead to any relaxation. To fix the energy and momentum conservation a third particle has to be involved which takes up excess energy in the relaxation process. For weak interactions, the system is therefore described by three body scattering processes \cite{Lunde2007}. 

In order to obtain theoretical estimates for the experimentally observed relaxation rates, we extend the three-body scattering formalism of Refs. \cite{Lunde2007,Karzig2010a} to higher energies and non-quadratic dispersions. To describe the time scales for decoherence due to inelastic processes we are interested in the out-scattering-rate of an hot electronic state (labeled by "1")
\begin{equation}
\Gamma=\sum_{k_{2}k_{3}q_{1}q_{2}q_{3}}\sum_{\sigma_{2}\sigma_{3}\sigma_{1'}\sigma_{2'}\sigma_{3'}}W_{123,1'2'3'}n_{2}n_{3}(1-n_{1'})(1-n_{2'})(1-n_{3'})\,,\label{eq:out}
\end{equation}
Following previous notations. Note that despite the strong spin orbit coupling of InAs the spin labels are still good quantum numbers in the absence of an external magnetic field. The momentum sums run over all distinct initial and final states, which can be achieved by restricting $k_1>k_2>k_3$ and $k_1'>k_2'>k_3'$ or by including appropriate prefactors that compensate for double counting of states. 

The three-body scattering matrix element $W_{123,1'2'3'}$ can be obtained from the generalized Fermi golden rule expression
\begin{equation}
W_{123,1'2'3'}=\frac{2\pi}{\hbar}|\langle1'2'3'|VG_{0}(\varepsilon_{i})V|123\rangle_{c}|^{2}\delta(\varepsilon_\mathrm{i}-\varepsilon_\mathrm{f})\,,\label{eq:threebody_wire}
\end{equation}
The subscript $c$ signals that only "connected" processes where all three particles contribute should be taken into account (thus excluding effective
two-particle processes). $G_{0}$
and $V$ are the free propagator and the interaction part of the Hamiltonian, respectively. They take the form
\begin{eqnarray}
G_{0}(\varepsilon_{i}) & = & \frac{1}{\varepsilon_{i}-H_{0}+{\rm i}0^{+}}\,,\\
H_0 & = & \sum_{k \sigma}\varepsilon(k)c^\dagger_{k\sigma}c_{k\sigma},\\
V & = & \frac{1}{2L}\sum_{k_{1}k_{2}q\sigma_{1}\sigma_{2}}V_{q}c_{k_{1}+q\sigma_{1}}^{\dagger}c_{k_{2}-q\sigma_{2}}^{\dagger}c_{k_{2}\sigma_{2}}c_{k_{1}\sigma_{1}}\,,
\end{eqnarray}
where $\varepsilon(k)$ describes the dispersion of the lowest subband, $L$ is the wire length and $V_q$ is the one dimensional Fourier transform of the Coulomb interaction. Using the above relations the three-body scattering amplitude can be decomposed into
\begin{equation}
\langle1'2'3'|VG_{0}V|123\rangle_{c}=\sum_{a'b'c'=P(1'2'3')}(-1)^{p}\delta_{\sigma_1, \sigma_{a'}}\delta_{\sigma_2, \sigma_{b'}}\delta_{\sigma_3, \sigma_{c'}}T_{a'b'c'}^{123}\,,
\end{equation}  
where $P(1'2'3')$ denotes all possible permutations of the primed labels and $p$ is the corresponding parity of the permutation. The decomposition allows to distinguish between the direct term that involves momentum transfers $q_i$
\begin{equation}
T_{1'2'3'}^{123}=\frac{1}{L^2}\sum_{abc=P(123)}\frac{V_{k_{a'}-k_a}V_{k_{c'}-k_c}\delta_{k_a+k_b+k_c,k_{a'}+k_{b'}+k_{c'}}}{\varepsilon(k_b)+\varepsilon(k_c)-\varepsilon(k_{c'})-\varepsilon(k_b+k_c-k_{c'})},
\end{equation}
and five exchange terms that arise from exchanging the final states. With the convention that particle 3 is a left mover while 1 and 2 are right movers there is only one forward-scattering exchange term $T^{123}_{2'1'3}$, while the remaining four exchange terms involve momentum transfers larger than $2k_F$. In this experiment we are working in the regime where $k_F>1/a$ (in terms of the wire radius $a$) which suppresses the back-scattering exchange terms. For large momenta of the injected hot electron also the forward-scattering exchange term (with momentum transfers of the order of $k_1-k_F$) will be suppressed.

Finally we require a concrete form of the Fourier transform of the Coulomb interaction $V_q$. Using the unscreened Coulomb interaction in real space $V(r)=e^2/4\pi \epsilon r$, where $\epsilon$ is the effective dielectric constant, the effective one dimensional Fourier transform takes the form 
\begin{equation}
V_q= \frac{e^2}{4\pi\epsilon}\int \frac{d {\bf q}_\perp}{(2\pi)^2} \frac{4\pi}{q_\perp^2+q^2}|F({\bf q_\perp})|^2,
\end{equation} 
with the form factor
\begin{equation}
F({\bf q_\perp})=\int d{\bf r_\perp} |\phi({\bf r_\perp})|^2 \text{e}^{-i {\bf q_\perp}\cdot{\bf r\perp}}\,,
\end{equation}
Here, we assumed a Gaussian wavefunction for the perpendicular confinement $\phi({\bf r_\perp})=\sqrt{2/\pi a^2}\exp[-(r_\perp/a)^2]$, which yields
\begin{equation}
V_q=-\frac{e^2}{4\pi\epsilon}\text{e}^{\frac{q^2a^2}{4}}\text{Ei}(-q^2 a^2/4)\,,
\end{equation}   
where Ei($x$) is the exponential integral function. As estimated in Sec. \ref{Theory} $V_q$ interpolates between $-\log(|q|a)$ for small $q\ll1/a$ and $1/(qa)^2$ for large $q\gg 1/a$.

With the above model, we perform the integrations in Eq.~\eqref{eq:out} numerically with no further approximations. using the dispersion and parameters fitted to the experiment (Appendix \ref{D1}). 

\subsection{\label{D3} Three-body inter-subbnad relaxation}

Although the non-monotonic behavior of the relaxation rate $\Gamma$ is an inherent property of a 1D non-linear band (Fig. \ref{Fig 4}b), to what degree it is pronounced depends both on the curvature of the dispersion and on the position of $\varepsilon_{F}$. If $\varepsilon_{F}$ is close to the bottom of the band, the phase space growth is limited by the finite depth of the Fermi sea, and this may contribute to saturation even before the momentum transfer is suppressed by the Coulomb interaction. Namely, the phase space will only grow linearly (rather than quadratically) for $q>k_{F}$ thus suppressing the growth of $\Gamma$ at $\varepsilon \sim \varepsilon_{F}$ . Moreover, smaller $\varepsilon_{F}$ results in smaller $v_{F}$, leading to a larger velocity mismatch between the hot and the excited electron, making the suppression of $\Gamma$ more pronounced. 

\begin{figure}
\includegraphics{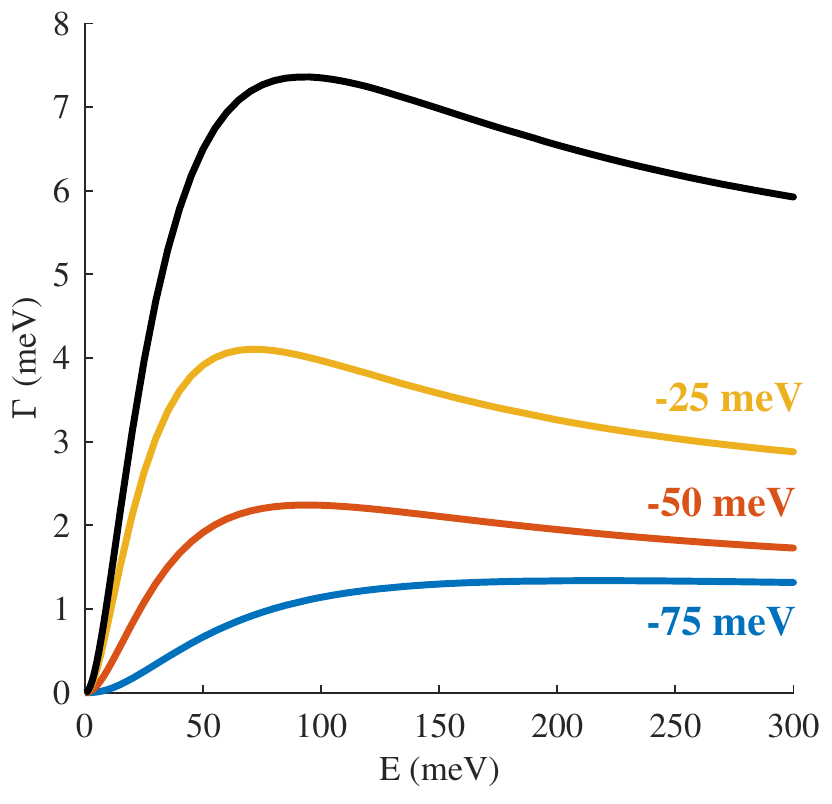} 
\caption{\label{Fig 12} Relaxation rate for individual sub-bands. Contributions to $\Gamma$ by exciting the Fermi sea of different subbands; the lowest subband (-75 meV below $\varepsilon_{F}$, in blue), two higher occupied subbands (-50 meV in red and -25 meV in yellow) and the sum of all three contributions (black). The energy denotes the bottom of the subbands, and we use energy spacing and dispersion extracted from the experiment. }
\end{figure}

In our experiment the relaxation of hot electrons from the lowest subband can be facilitated by excitations of particle-hole pairs from higher occupied subbands. Such subbands, having smaller $\varepsilon_{F}$ and $v_{F}$, show a more pronounced non-monotonicity while having a large density of states and therefore contribute significantly to the overall trend captured by summing over all subbands. Fig. \ref{Fig 12} shows the contributions to $\Gamma$  by exciting quasiparticles of each of the occupied subbands individually, with energy spacing and dispersion which was inferred from our experiment. Their sum (black curve) was used for the theoretical model shown in Figs. \ref{Fig 2}c and \ref{Fig 3}e.  

\section{Data analysis}

\subsection{\label{E1} Extracting the phase coherence length from the QPI patterns}

The data presented on Fig. \ref{Fig 2}a is obtained by filtering all the non-dispersing features present in Fig. \ref{Fig 1}d. The analysis is based on the dispersion, $E\left(q\right)$, which we extract from Fourier analysis (Fig. \ref{Fig 1}e). At each energy we perform a ''spatial lock-in'' analysis by cross-correlating the LDOS, normalized by its mean value away from the nanowire end, with a single period of a cosine wave of the corresponding momentum transfer $q_{E}$ extracted from the dispersion. 

\begin{equation}
\operatorname{XLDOS} (E,x) = \int {\operatorname{LDOS} (E,x + x')\operatorname{K} (E,x')dx'}, 
\end{equation}

\begin{figure}
\includegraphics{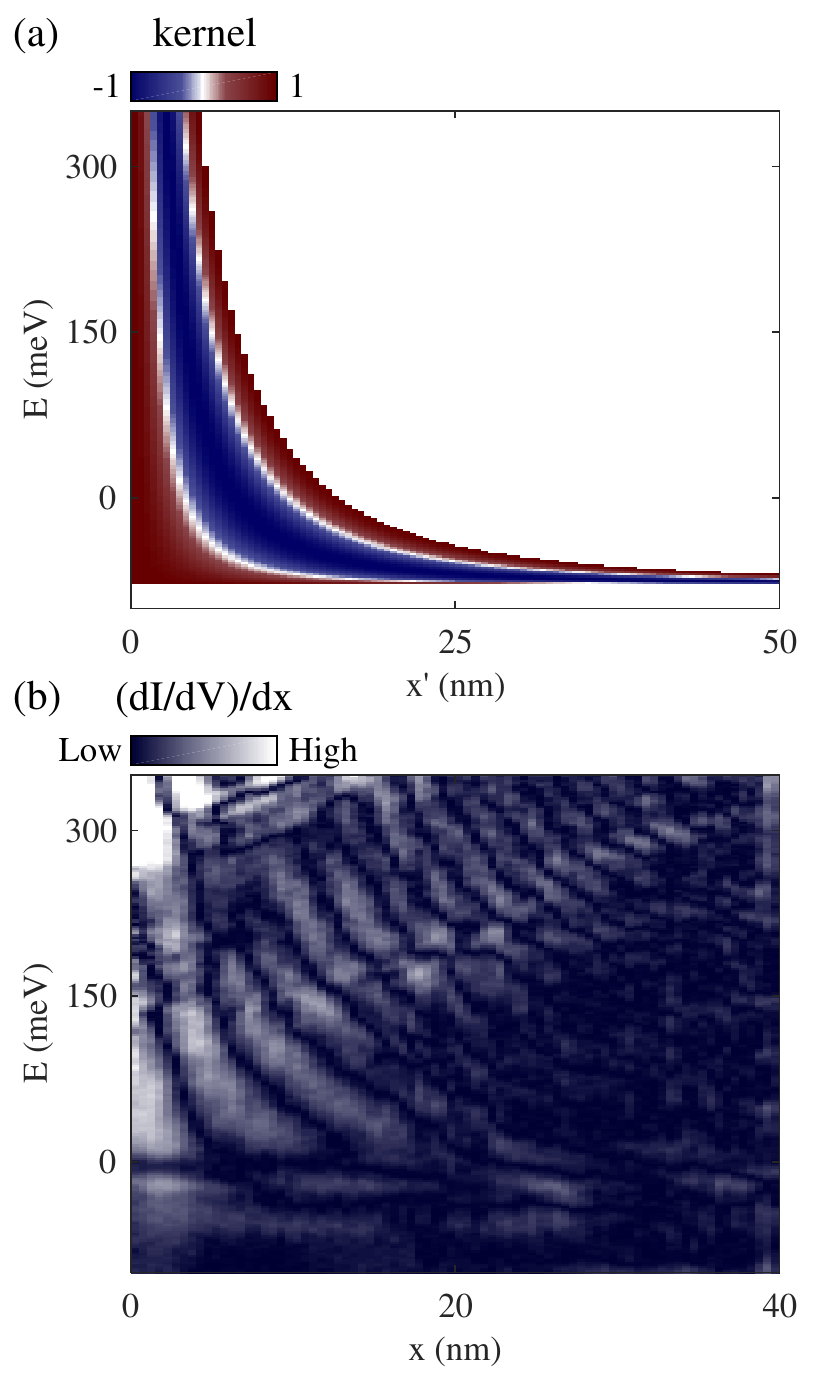} 
\caption{\label{Fig 13} (a) The kernel used to calculate the cross correlation: $\cos\left(q_{E} x' \right) \mathrm{rect}\left(q_{E} x' \right)$ is shown for each energy $E$ as a function of $x'$. This is simply a single period of a cosine with a wavelength corresponding to the inferred momentum transfer $q_{E}$. (b) Spatial derivative of the line-cut shown in Fig. \ref{Fig 1}d. The profile of the non-monotonic decay of the standing wave pattern can easily be traced. }
\end{figure}

Where the kernel of the correlation is a single period of the wavelength corresponding to the energy:

\begin{equation}
\mathrm{K(E,x)} =
\begin{cases}
cos \left( {q_{E}x} \right) & \text{$0<x<2\pi/q_{E}$} \\
0 & \text{otherwise},
\end{cases}
\end{equation}

as shown in Fig. \ref{Fig 13}a. The features that do not have the same spatial periodicity as the standing wave at that energy are eliminated due to the orthogonality of the Fourier series. Finally, the data shown in Fig. \ref{Fig 2}a-b is the absolute value of the cross-correlated LDOS. 

In Fig. \ref{Fig 13}b we demonstrate that the non-monotonic behavior is clearly captured even by performing the simpler operation of numerical spatial differentiation $\Delta \left(dI/dV\right)/\Delta x$ . The differentiation eliminates all features that do not disperse with energy such as Van-Hove singularities, and therefore enhances the relative contribution of the standing wave. The main disadvantage of differentiation though, is that it complicates the functional form of the decaying wave $d\left(\sin\left(q x\right) \mathrm{e}^{-2x/L_{\varphi}}\right)/dx$.  

\subsection{\label{E2} Extracting the energy broadening of the resonances}

\begin{figure}
\includegraphics{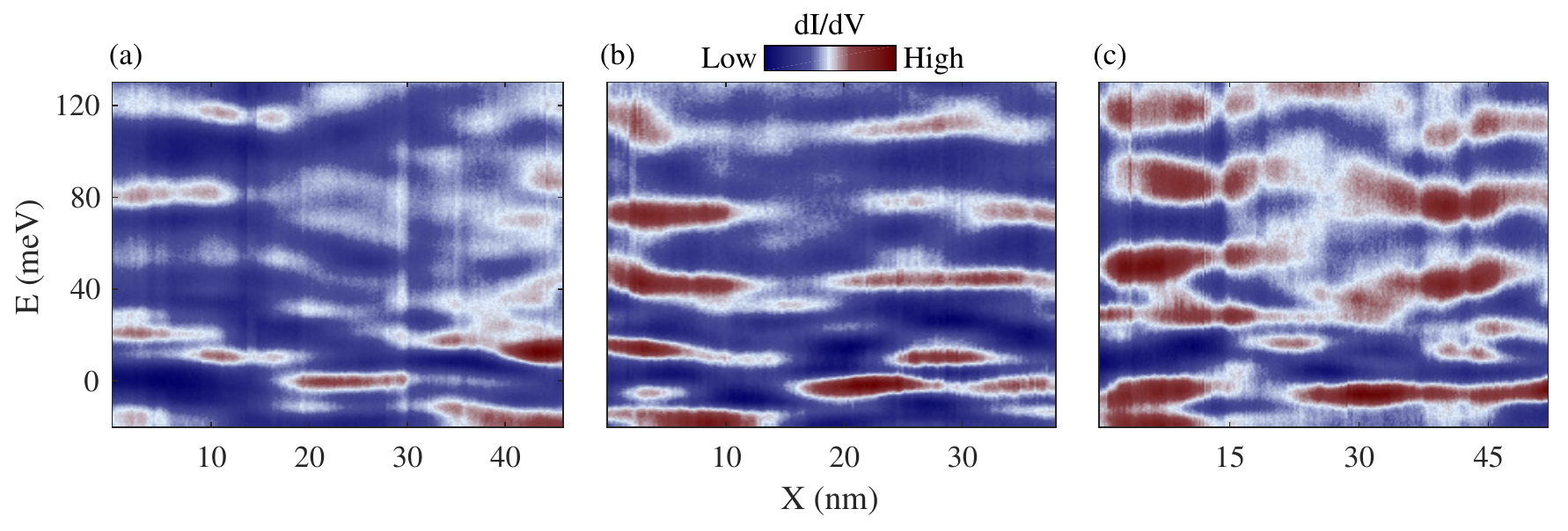} 
\caption{\label{Fig 14} . Quantized resonances in different terraces. (a), (b), and (c) show the dI/dV maps from three different terraces, averaged over 12 linecuts across the nanowire ($Y$). A smooth background has been subtracted for visibility. The observed puddels are the quantized resonances used to determine the lifetime of the states. }
\end{figure}

We studied the resonances generated by scattering from adjacent stacking faults (Fig. \ref{Fig 3}). The stacking faults backscatter the electrons with a certain probability $R$ giving rise to the quantization seen in the spectrum. Their spatial distribution along the nanowire axis ($X$) is rather symmetric around the center of the terrace (Fig. \ref{Fig 14}), indicative of confinement of the electronic wavefunction. By fitting the energy broadening of these resonances we deduced the reflectivity of electrons through stacking faults, the lifetime of the electronic states and their relaxation properties.

\begin{figure}
\includegraphics{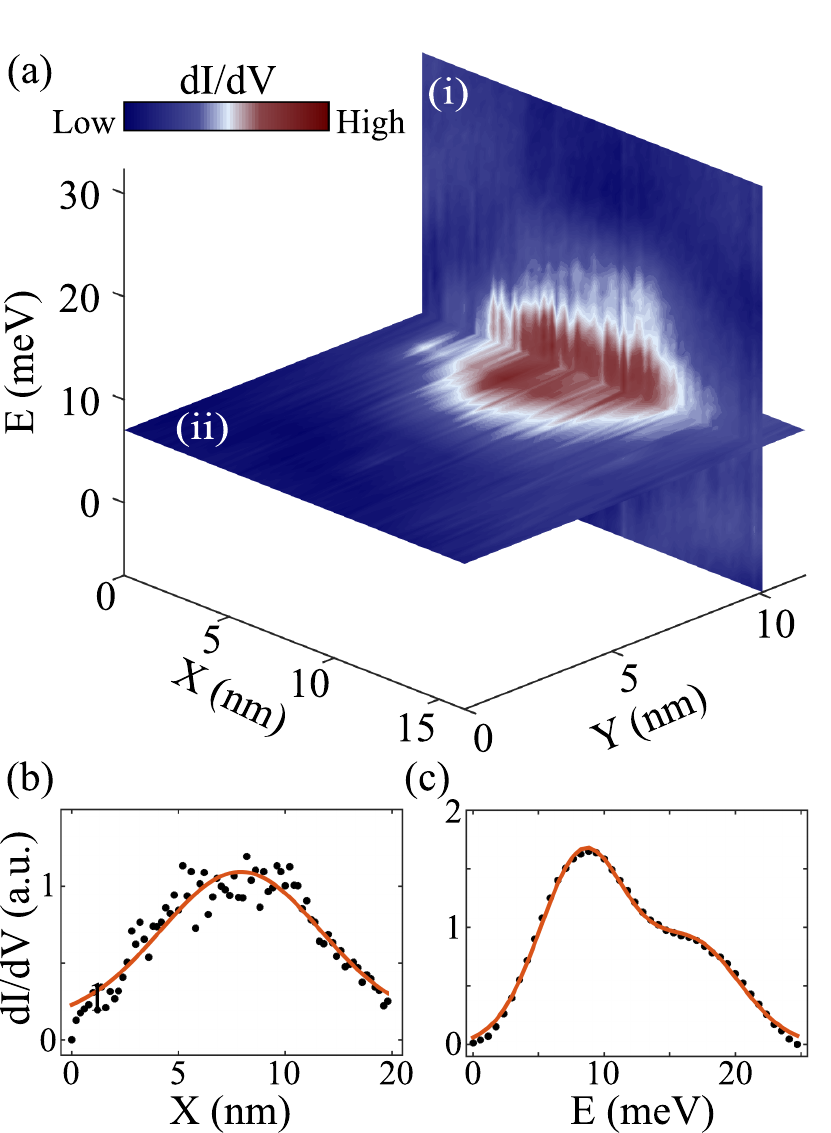} 
\caption{\label{Fig 15} Measuring the energy broadening of a resonance peak. (a) 3D slices of a resonance in the dI/dV map, showing its energy (i) and spatial (ii) distributions. (b) Gaussian fitting of the spatial distribution of the peak at a given linecut ($Y_{0}$) and energy. (c) Multiple gaussian model fitting of the energy distribution of the peak, determining its width and peak energy. }
\end{figure}

The resonances are spatially spread both along the nanowire axis ($X$) and across it ($Y$), calling for a careful analysis to ensure a proper determination of their accurate energy broadening $\Gamma$. The procedure of this analysis is demonstrated in Fig. \ref{Fig 15}. For each resonance we first determine the position across the nanowire $Y_{0}$, at which it is broadest in energy as shown in Fig. \ref{Fig 15}a. Next, to determine its peak position along the nanowire $X_{0}$, we average its dI/dV over a narrow (one-$\sigma$) energy window around its peak energy, shown in Fig. \ref{Fig 15}b, and extract its peak position by fitting a gaussian. Finally, we plot the dI/dV energy profile at the peak position ($X_{0}, Y_{0}$) and fit a single (or double) gaussian profile whose mean and variance are the resonance peak energy and width $\Gamma\left(E\right)$, plotted in Fig. \ref{Fig 3}e.

\subsection{\label{E3} Extracting the stacking faults reflectivity from resonance broadening}

To relate resonance broadening to the reflection coefficient of electrons scattered by adjacent stacking faults we employed Fabry-P\`{e}rot analysis similar to ref. \cite{Seo2010}. Calculating the wavefunction amplitude resulting from a superposition of all scattering paths in the single band case, gives the expression for the quantized LDOS.

\begin{equation}
\mathrm{LDOS}\left(E,x\right) = \frac{1}{N} \left( \frac{4+4r^{2}\cos^{2}\left(2k_{E}x\right) + 8\cos\left(2k_{E}x\right)\cos\left(\theta_{r}+2k_{E}\left(L/2\right)\right)}{1+r^{4}-2r^{2}\cos\left(2\theta_{r}+2k_{E}L\right)} \right)
\end{equation}    

\begin{figure}
\includegraphics{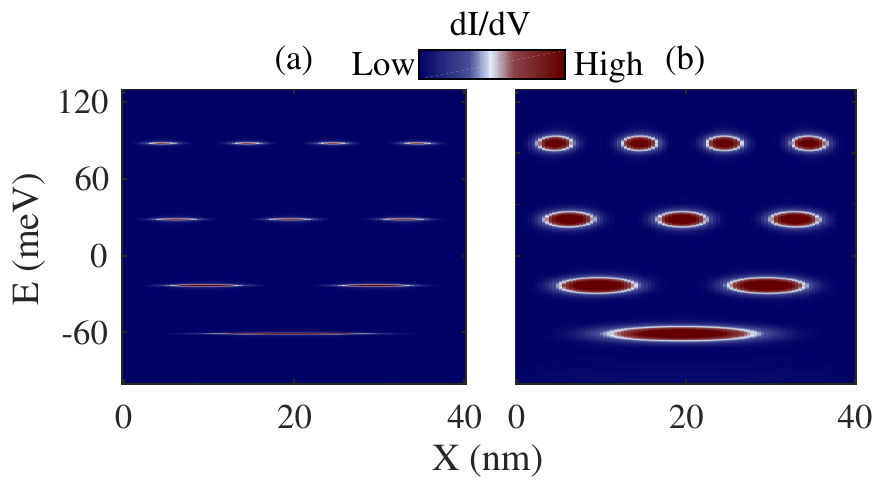} 
\caption{\label{Fig 16} Quantized resonances in a confined 1D system. Calculated LDOS in a quantum box with $\left|R\right|^{2} = 0.9$ . (a) Resonances of the lowest energy band, calculated from the experimentally inferred dispersion.  (b) Same resonances after instrumental broadening due to temperature ($\sim1$ meV) and finite probing amplitude (3 meV). }
\end{figure}

Here, $R=r\mathrm{e}^{i\theta}$ is the reflection coefficient, $N$ is a normalization factor, $L$ is the length of the terrace and $k_{E}$ is the wavevector at energy $E$ inferred from the experimentally measured dispersion. The resulting LDOS (for $\left|R\right|^{2} = 0.9$  as inferred from the experiment) is plotted in Fig. \ref{Fig 16}a, and shows the expected alternating symmetric and antisymmetric modes. A resonance appears when the denominator is minimized, namely when $\cos\left(2\theta_{r}+2k_{E}L\right) = 1$. Expanding the denominator gives the full-width half-maximum (FWHM) of each energy resonance.

%





\bibliography{Refs}

\end{document}